\documentclass[12pt,preprint]{aastex}

\newcommand{\teff}{$T_{\rm eff}$}

\newcommand{\Mch}{$M_{\rm CH}$}
\newcommand{\chandra}{{\it Chandra}}
\newcommand{\newton}{{\it Newton}}

\shorttitle{X-Ray Spectroscopy of CAL 83}
\shortauthors{Lanz et al.}

\begin{document}

\title{NLTE Model Atmosphere Analysis of the LMC Supersoft X-ray Source CAL 83}


\author{Thierry Lanz\altaffilmark{1,2}, Gisela A. Telis\altaffilmark{2},
        Marc Audard\altaffilmark{2}, Frits Paerels\altaffilmark{2}, \\
        Andrew P. Rasmussen\altaffilmark{2}, and Ivan Hubeny\altaffilmark{3}}

\altaffiltext{1}{Department of Astronomy, University of Maryland,
                   College Park, MD 20742}
\altaffiltext{2}{Columbia Astrophysics Laboratory, Columbia University, 550 West 120th Street,
                   New York, NY 10027}
\altaffiltext{3}{Steward Observatory, University of Arizona, 933 N Cherry Avenue, Tucson, AZ 85721}

\email{tlanz@umd.edu, gat9@columbia.edu, audard@astro.columbia.edu, frits@astro.columbia.edu,
       arasmus@astro.columbia.edu, hubeny@as.arizona.edu}

\setcounter{footnote}{3}

\begin{abstract}
We present a non-LTE model atmosphere analysis of \chandra\  HRC-S/LETG 
and XMM-\newton\  RGS spectroscopy of the prototypical supersoft source
CAL~83 in the Large Magellanic Cloud. Taken with a 16-month interval,
the \chandra\  and XMM-\newton\  spectra are very similar. They reveal a
very rich absorption line spectrum from the hot white dwarf photosphere,
but no spectral signatures of a wind.
We also report a third X-ray off-state during a later \chandra\  observation,
demonstrating the recurrent nature of CAL~83. Moreover, we found evidence of
short-timescale variability in the soft X-ray spectrum. We completed the analysis
of the LETG and RGS spectra of CAL~83 with new NLTE line-blanketed model atmospheres
that explicitly include 74 ions of the 11 most abundant species. We successfully
matched the \chandra\ and XMM-\newton\  spectra assuming a model composition with LMC metallicity. 
We derived the basic stellar parameters of the hot white dwarf, but the current state
of atomic data in the soft X-ray domain precludes a detailed chemical analysis. 
We have obtained the first direct spectroscopic evidence that the white dwarf is
massive ($M_{\rm WD} \ga 1 M_\odot$). The short timescale of the X-ray off-states
is consistent with a high white dwarf mass. Our analysis thus provides direct support
for supersoft sources as likely progenitors of SN~Ia.
\end{abstract}



\section{Introduction}
\label{Intro}

Because of the overall importance of Supernovae type Ia (SNe Ia) in
astrophysics and in cosmology, the identification of their progenitors is
a pressing issue. Theoretical models suggest that SNe~Ia arise from the
thermonuclear explosion of a carbon-oxygen (CO) white dwarf (WD) that
has grown to the Chandrasekhar mass, \Mch\  \citep{hoyle60, arnett69}.
To date, the most promising formation channel resulting in SNe~Ia
involves accreting WD's that sustain steady nuclear burning close to
their surface.  When they form, CO WD's have masses between 0.7 and
1.2\,$M_\sun$ depending on the star initial mass \citep{weidemann87}.
SN~Ia progenitors, therefore, must be in close
binaries where the WD can gain several tenths of a solar
mass donated by a companion to reach \Mch. Because we lack a direct
determination of the progenitor properties, we have to rely on indirect
arguments to determine their nature. 

Two main scenarios have been proposed, involving either the merger of two WD's
\citep[the double-degenerate (DD) scenario;][]{iben84, webbink84},
or a single WD accreting from a normal companion
\citep[the single-degenerate (SD) scenario;][]{whelan73}.
The DD scenario provides a natural way to explain the absence
of hydrogen lines in SN~Ia spectra, but hardly explains why SNe~Ia could
be ``standard candles'' since the explosion should depend on the respective
properties of the two WD's. Moreover, while the DD scenario provides an obvious
way to form a more massive object, possibly reaching \Mch, the few double WD systems
identified to date that could merge in a Hubble time have a total mass
smaller than \Mch\ \citep{saffer98}. New large-scale searches for DD systems,
however, have recently yielded promising results \citep{napiwotzki02, napiwotzki03}, and
good candidates might be discovered soon. Finally, the merger scenario has been
suggested to lead to an accretion-induced collapse rather than 
to a SN~Ia event \citep{nomoto85, saio85}.

The SD scenario is thus generally favored today. The principal issue of
the SD scenario is, however, to ascertain if the WD can at all accrete
sufficient mass to lead to a SN~Ia event. This topic
has been hotly debated in the last 10 years to sort out the competing effects
of mass ejection and accretion. \cite{nomoto79} first pointed out that
an accreting WD would quickly ignite a small amount of accreted material in
a hydrogen shell-burning and undergo a weak shell flash. Most of the accreted
mass then accumulates in an extended, supergiant-like envelope that might
be lost during a subsequent common-envelope phase. \cite{hachisu96, hachisu99}
argued that starting at a certain critical accretion rate a strong wind 
develops, regulating the mass transfer between the mass-losing star
and the accreting WD. The WD accretes at most at the critical rate,
$\dot{M}_{\rm cr} = 9.0\,\times\,10^{-7}\,(M_{\rm WD}/M_\odot - 0.50)\,M_\odot$\,yr$^{-1}$,
and the rest is blown off in the wind. At a later stage the accretion rate drops,
and the strong wind stops; the final outcome depends on the mass
accreted before this point, because stronger flashes will develop at
lower accretion rates. \cite{cassisi98} made the opposite argument. The basic
physical processes are the same, but they argued that dynamical
helium-burning flashes probably would hamper accretion
and, therefore, the WD most likely would not reach \Mch. The most recent
theoretical calculations support the idea that the accreting WD may
eventually reach \Mch. \cite{yoon03} have built the first detailed
binary star evolution model where the WD gains enough mass from a
$1.6\,M_\odot$ helium star. This model involves higher accretion rates
(few $10^{-6}\,M_\odot$\,yr$^{-1}$). Subsequently, \citet{yoon04} have
shown that rotation tends to stabilize the helium-burning shell, thus
increasing the likelihood that accreting WD would reach the stage of
central carbon ignition.

Owing to the difficulties met by theoretical works, an observational approach
is essential to characterize SN~Ia progenitors. The most promising candidates
are Close Binary Supersoft X-ray Sources (CBSS) that were revealed by ROSAT
as a new class of close binaries with ultrasoft X-ray spectra showing no
emission above $0.5 - 1$\,keV \citep{trumper91}. \citet[][vdH92]{vdH92} argued that
such a soft emission is not consistent with accreting neutron stars or
black holes, which show emission peaking at $1 - 10$\,keV. They proposed the now
classical model of an accreting WD, where a relatively high-mass WD
$(0.7 - 1.2\,M_\odot)$ sustains steady burning of the hydrogen-rich accreted
material. They argued that accretion has to occur at a finely-tuned rate, 
$1.0\,\times\,10^{-7}\la\dot{M}\la 4.0\,\times\,10^{-7}\,M_\odot$\,yr$^{-1}$. At lower
rates, hydrogen burning is unstable and occurs in flashes, while an extended
envelope forms at higher rates. The stellar luminosity is then dominated by
hydrogen burning which liberates an order of magnitude or more of energy than
accretion itself. Typical temperatures, $kT\approx 30 - 80$\,eV, are derived from
blackbody simulations and WD model atmospheres after correction for interstellar
extinction \citep{kahabka97}. Based on the vdH92 model, \cite{rappaport94} discussed
the formation and evolution of CBSS, reproducing their typical luminosities,
effective temperatures and orbital periods. They estimated that the rate of
Galactic SNe~Ia associated with the evolution of CBSS might reach 0.006~yr$^{-1}$.
Most recently, \cite{ivanova04} extended the vdH92 work, developing a semianalytical
model to investigate the evolution of binaries consisting of main-sequence stars
with WD companions in the thermal mass-transfer phase. They accounted for the
stabilizing effect of the WD wind, and characterized the different conditions
leading to different outcomes, double WD's, (sub-)\Mch\  SNe~Ia, or
accretion-induced collapse. To evolve towards a SN~Ia explosion, \cite{ivanova04}
argued that the WD and the donor star should be initially relatively massive,
$M_{\rm WD}\ga 0.8\,M_\odot$ and $M_{\rm d}\ga 2\,M_\odot$, with a mass ratio
$M_{\rm d}/M_{\rm WD}$ smaller than about 3.

The SD scenario recently gained further observational support. \cite{hamuy03}
reported the detection of a narrow H$\alpha$ emission line in SN2002ic, providing
the first evidence of hydrogen-rich circumstellar material associated with a SN~Ia.
This detection might be the missing smoking gun for the SD scenario, though
\cite{livio03} argued that this conclusion might be premature based on the absence
of a detection in all other SN Ia spectra.

The exact nature of SN~Ia progenitors thus remains an open problem.  Determining
the properties of CBSS may, therefore, provide the strongest empirical case supporting
the SD scenario. \chandra\  and XMM-\newton 's capabilities in obtaining high-resolution
spectra in the soft X-ray domain now open the possibility to determine spectroscopically
the stellar parameters of the WD's in CBSS, in particular the surface gravity and
hence the WD mass. We present in this paper a non-LTE (NLTE) analysis of the \chandra\  and
XMM-\newton\  spectra of the prototypical CBSS CAL~83. 
We start by summarizing previous results on CAL~83 in \S2, and the
new observations are discussed in \S3 and \S4.
The NLTE model atmospheres are detailed in \S5. Sect.~6 presents the spectrum analysis,
leading to a conclusion that the WD in CAL~83 is massive (\S7).


\section{The CBSS CAL 83}
\label{CAL83}

The {\em Einstein} observatory survey of the Large Magellanic Cloud (LMC)
revealed two sources with an ultrasoft spectrum, CAL~83 and CAL~87
\citep{long81}. CAL~83 (=~RX~J0543.5-6823) was identified with a variable, blue,
$V\approx 17$, point-like source with an orbital period of 1.04 days
\citep{cowley84, smale88}. Soft X-ray spectra of CAL~83 have been subsequently
obtained with ROSAT PSPC \citep{greiner91} and {\em Beppo}SAX LECS \citep{parmar98}.
Because of the limited spectral resolution of these observations, no spectral
features were visible. The spectral energy distribution was modeled using first
blackbodies and then WD model atmospheres. A blackbody analysis of the ROSAT
observations implies the puzzling result that CBSS have radii typical
of WD's but radiate at or above the Eddington limit \citep{greiner91}.
\cite{heise94} constructed the first LTE model atmospheres of CBSS. They
showed that these models predict a higher flux in the soft X-ray range, and
hence do not require super-Eddington luminosities to fit the ROSAT data.
\cite{hartmann97} extended Heise et~al.'s work, investigating the importance
of departures from LTE. NLTE effects are expected in high gravity
objects if they are sufficiently hot \citep{dreizler93, NLTE2}. Indeed,
Hartmann~\& Heise found significant differences between LTE and NLTE model
spectra in the temperature range of WD's in CBSS.
The main effect consists in the overionization of heavy species. NLTE model
atmospheres and NLTE effects are further discussed in \S5 and \S6.

Significant advances are now expected from \chandra\  and XMM-\newton\ spectrometers
because of their higher spectral resolution. XMM-\newton\  RGS spectroscopy of CAL~83 was
first obtained aiming at deriving the fundamental stellar and binary
parameters \citep{paerels01}. The spectrum shows a very rich line structure.
The application of NLTE model atmospheres to analyze the RGS spectrum was,
however, only marginally successful and no satisfactory, detailed match to
the RGS data was achieved. This paper takes over from the Paerels et~al. paper
and includes new \chandra\  LETG spectroscopy extending
beyond the RGS cutoff at 40\,\AA\  as well. This extension to lower energies provides
crucial new data to normalize the NLTE model spectra and, hence, to determine
the parameters of CAL~83.


\section{Observations}
\label{Observ}

CAL~83 was observed in April 2000 by XMM-\newton\  and at 3 epochs (November 1999,
August 2001, October 2001) by \chandra, see Table~\ref{LogTbl} for a log of the
observations. In November 1999 and in October 2001, CAL~83 was found to be in a
X-ray off-state (see \S\ref{XVar} and Greiner~\& Di Stefano 2002). The XMM-\newton\  RGS
observation was described by \cite{paerels01}. We detail here only the August 2001
\chandra\  observation.

On 2001 August 15, we observed CAL~83 for 35.4~ksec with \chandra, using the
High Resolution Camera (HRC-S) and the Low Energy Transmission Grating (LETG).
This setup affords a spectral coverage between 1 and 175\,\AA\  at a nominal resolution
of 0.05\,\AA\  and, therefore, provides full coverage from the Wien tail of the
energy distribution to the cut-off due to interstellar extinction. 
The data were processed with CIAO, version 3.0.2\footnote{http://cxc.harvard.edu/ciao/}.
We applied the observation-specific status and Good Time Interval 
filters, as well as an additional, non-standard filter on the pulse-height value as a function
of dispersed-photon wavelength. Wargelin~\&
Ratzlaff\footnote{http://cxc.harvard.edu/cal/Letg/Hrc\_bg/}
suggested to use a light pulse-height filter 
which can reduce the background rate in the HRC-S/LETG configuration by 50-70\%
with minimal X-ray losses and virtually no possibility of introducing spurious spectral
features. To test the effects of such
a filter on our data, we simultaneously completed the data
reduction with and without the filter. After a comparison of the output event lists and spectra
confirmed that no significant changes had been made to the valid photon events, we retained 
the pulse-height filter and used the filtered data for the remainder of the analysis.


We used the canned, on-axis first-order grating redistribution matrix file (gRMF) for the
LETG spectrum, but we ran the CIAO task ``fullgarf'' to generate the grating ancillary response
files (gARFs) for the positive and negative first spectral order. Furthermore, we multiplied
the gARFs by the provided encircled energy auxiliary response file in order to correct for
the efficiency of the LETG ``bow-tie'' extraction region. The positive and negative order
spectra were co-added to increase the signal-to-noise ratio. We merged the corresponding
gARFs. Note that we did not correct for the effect of higher spectral orders since they are
negligible in the $20-70$~\AA\  faint spectrum of CAL 83.

\begin{figure}
\epsscale{.99}
\plotone{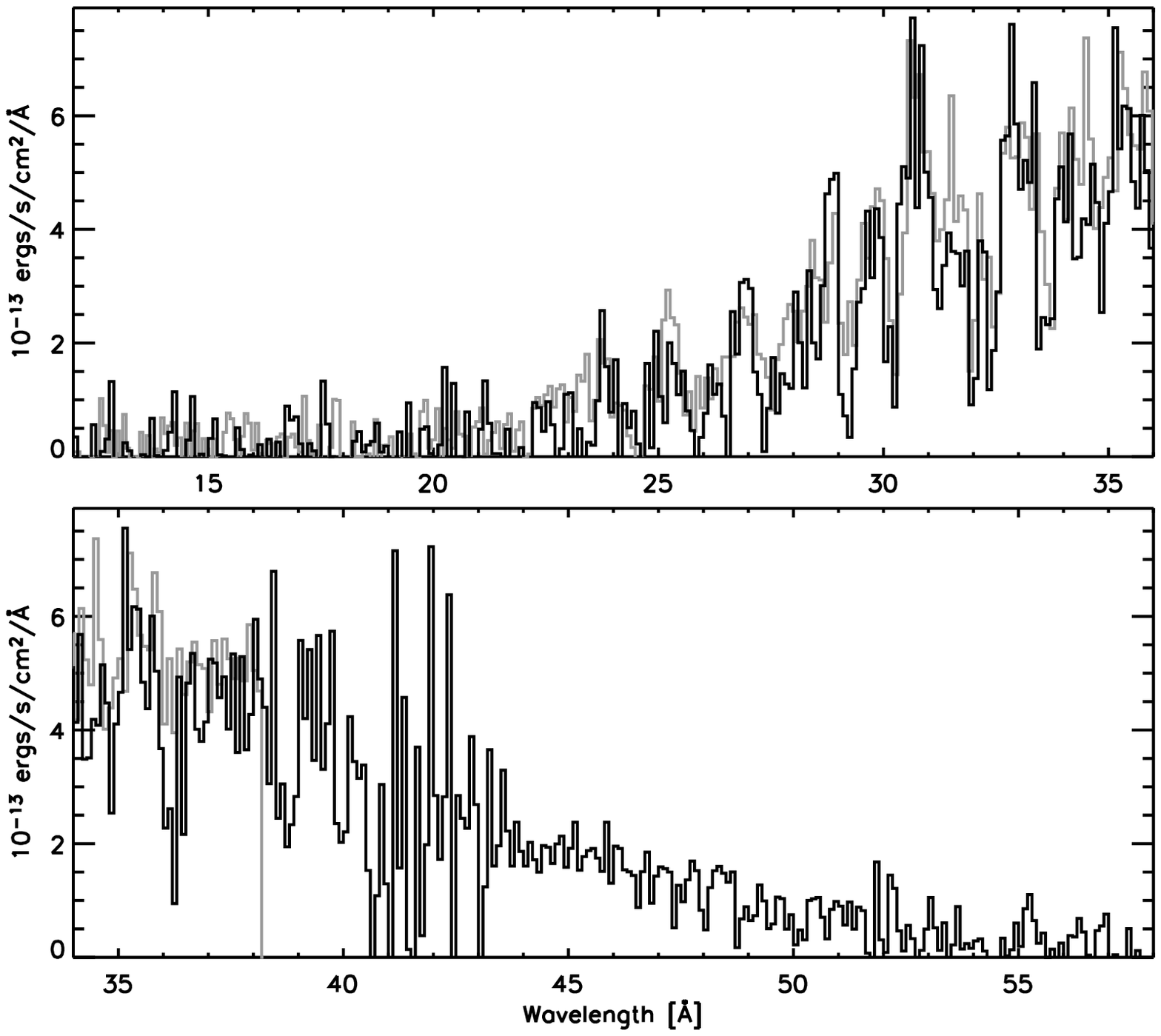}
\figurenum{1}
\caption{\chandra\  LETG (black line) and XMM-\newton\  RGS (grey line) spectra of CAL~83. 
The spectra were rebinned by a factor of 8.
\label{LETGFig}}
\end{figure}

\begin{figure}
\epsscale{.99}
\plotone{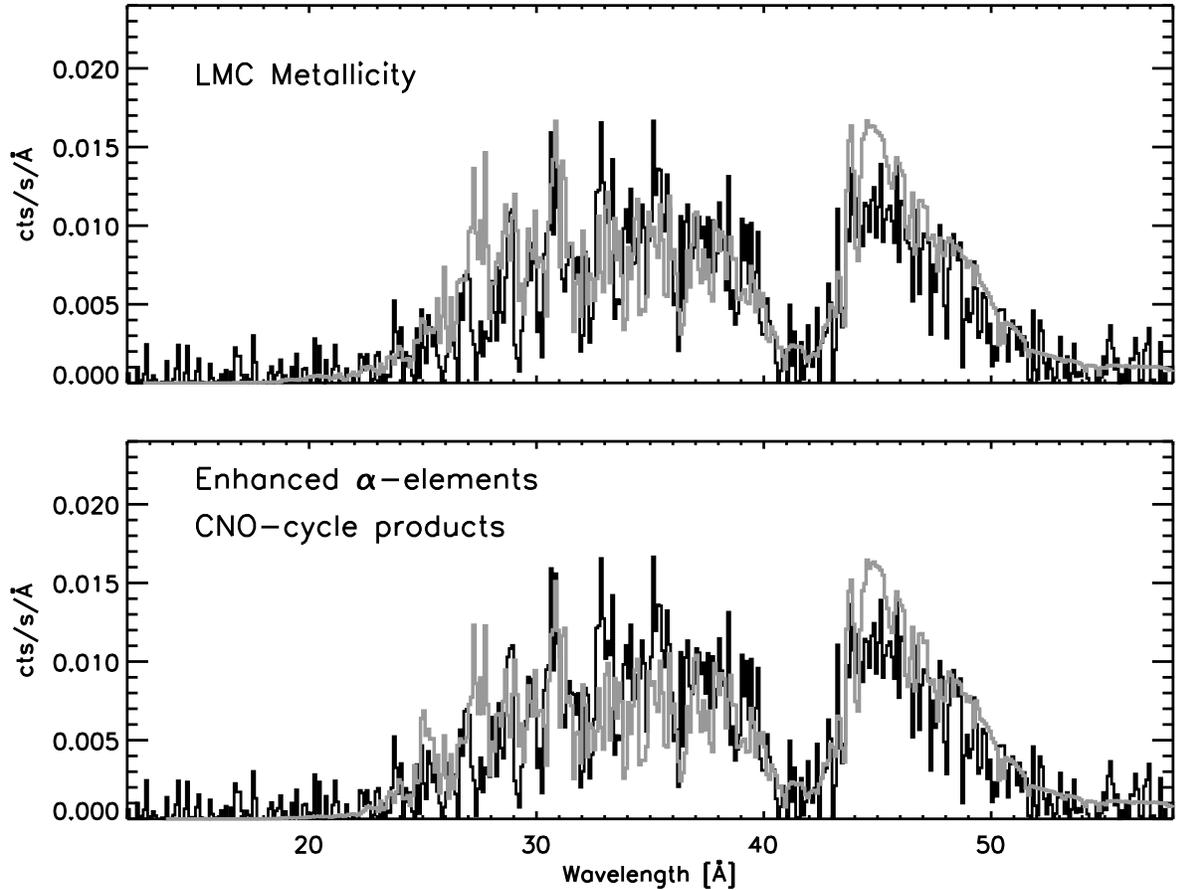}
\figurenum{7}
\caption{Best model fit to the LETG spectrum, \teff = 550,000~K, $\log g = 8.5$,
         assuming two different surface compositions.
         Top: LMC metallicity (half solar); bottom: LMC metallicity with enhanced
         $\alpha$-element abundances and CNO abundances typical of CNO-cycle processed material. 
\label{ChemFig}}
\end{figure}

Fig.~\ref{LETGFig} compares the RGS spectrum (we only use RGS1 data) and the LETG spectrum,
calibrated in flux with the instrumental response file.
We chose to rebin the spectra by a factor of 8, as the best compromise 
between signal-to-noise ratio and resolution, thus resulting in a resolution slightly
lower than the nominal resolution. The two spectra are very similar, both
in term of the spectral energy distribution and of the line features. The LETG spectrum is very
noisy between 40 and 44\,\AA, because of the very low effective area due to instrumental
(HRC-S) absorption by the carbon K edge at 43.6\,\AA. This instrumental absorption is clearly
seen in Figs.~\ref{MFitFig}-\ref{ChemFig} and is well represented in the response
function applied to the NLTE model spectra.


\section{X-ray Variability}
\label{XVar}

Although the main scope of the \chandra\  and XMM-\newton\  observations were
to obtain the first high-resolution X-ray spectra of CAL~83, we have extended
our analysis to look for X-ray flux variations during the observations.

\subsection{A New X-Ray Off-State}
\label{sect:offstate}


X-ray off-states have been rarely caught, with two exceptions in April 1996 
by ROSAT \citep{kahabka96} and in November 1999 by \chandra\ 
\citep{greiner02}. We report here another X-ray off-state. 
Because of solar flare activity, our \chandra\  HRC-S/LETG
observation in August 2001 was interrupted after 35~ksec and rescheduled for
completion in October 2001. This latter exposure, however, did not detect X-rays
from CAL~83 in 62~ksec, indicating a X-ray off-state.
We extracted counts at the expected position (obtained from
the August \chandra\ LETG observation which detected CAL~83) using a circle of 1\farcs 4
radius (95\% encircled energy for the mirror point-spread function, PSF),
and extracted background counts from a concentric annulus having an area 50 times larger.
We obtained 34 and 1223 counts, respectively, for a 61.55~ksec exposure. We then followed
the approach of \citet{kraft91} to determine the upper confidence limit using
a Bayesian confidence level of 95\%. This upper limit is 21.045 counts. 
Our adopted unabsorbed model spectrum (see \S\ref{Resu}) corresponds to
$L_\mathrm{X} = 2.7 \times 10^{37}$~ergs~s$^{-1}$
($0.1 - 10$~keV), i.e., a zeroth order count rate of 0.10 ct~s$^{-1}$ (i.e. an absorbed
flux at Earth of $8.3 \times 10^{-12}$~ergs~s$^{-1}$~cm$^{-2}$).
Consequently, the
X-ray off-state corresponds to an upper limit of $L_\mathrm{X} < 9.2 \times 10^{34}$~ergs~s$^{-1}$
(the upper limit of the absorbed flux at Earth is
$F_\mathrm{X} < 2.8 \times 10^{-14}$~ergs~s$^{-1}$~cm$^{-2}$).
The upper limit on $L_\mathrm{X}$ is similar to that reported by \citet{greiner02} during
the 1999 X-ray off-state. Assuming a constant bolometric luminosity, they have then infered
a temperature limit, $kT\la 15$\,eV, that agrees well with our own estimate based on NLTE model
atmospheres.

Two important caveats need to be mentioned, however. First, the extraction radius
probably underestimates the encircled energy since the zeroth order PSF
with LETG is not exactly similar to the mirror PSF. Thus, in principle,
a larger extraction region (e.g., $3\arcsec$) should contain an encircled
energy fraction closer to 100\%. Since the zeroth order effective area is
provided for a fraction of 100\%, such a larger radius could, therefore, provide a
more accurate estimate of the upper limit. Nevertheless, the second issue, that is 
the assumption of a similar model during the on-state and the off-state
for estimating the count-to-energy conversion factor, affects the determination
of the upper limit much more severely. \cite{greiner02} discussed several mechanisms
to explain the relation between X-ray off-states and the optical variability. Since
they were not able to reach a definitive conclusion, we do not know for instance
if the spectrum shifts to lower energies, which thus leaves a serious uncertainty
on the way of establishing an upper limit to the X-ray luminosity during the off-states.
The calculated upper limit, therefore, remains indicative only.

The original 1996 off-state was interpreted in two contrasted ways. On one hand, \cite{alcock97}
considered a model of cessation of the steady nuclear burning related to a drop in
the accretion rate. The WD would need to cool quickly to explain the observed off-state,
and the timescale of the off-state implies a massive WD.
This model was based on the combination of optical and X-ray variability,
and was later criticized by \citet{greiner02} who found that optical low states were
delayed by about 50 days relative to the X-ray off-states. On the other hand, \cite{kahabka98}
argued that the off-state is caused by {\em increased} accretion resulting in the swelling
and cooling of the WD atmosphere. Assuming adiabatic expansion at constant luminosity
and a realistic accretion rate ($\dot{M}\approx 10^{-6}~M_\odot\,{\rm yr}^{-1}$), he argued
from the characteristic timescale of the off-state that the WD is massive,
$M_{\rm WD}\ga 1.2 M_\odot$.

\citet{greiner02} discussed extensively the X-ray off-states and optical variability
of CAL~83. They believe that a cessation of nuclear burning was highly unlikely because the
short turnoff time requires the WD to have a mass close to \Mch, while earlier spectrum
analyses (e.g., Parmar et~al. 1998) suggested that the WD mass was not so high. Assuming then
that the luminosity stays roughly the same, Greiner~\& Di Stefano investigated two ways
to explain the off-states, namely an increase of the photospheric radius or absorption
by circumstellar material that would in both cases shift the emission to another spectral
domain. From a careful study of the variation patterns, they raised a number of issues for
the two models indicating that the optical variability could not be understood with a simple
model but requires some complex interaction between the photospheric expansion and the disk. 

CAL~83 was considered as the prototypical CBSS undergoing steady nuclear burning. However,
the observation of a third off-state now demonstrates the recurrent nature of the phenomenon
in CAL~83. We defer further discussion on the origin of the off-states after the spectrum
analysis.

\subsection{Long-Term and Short-Term Time Variability}
\label{sect:timevar}

\begin{figure}
\epsscale{.99}
\plotone{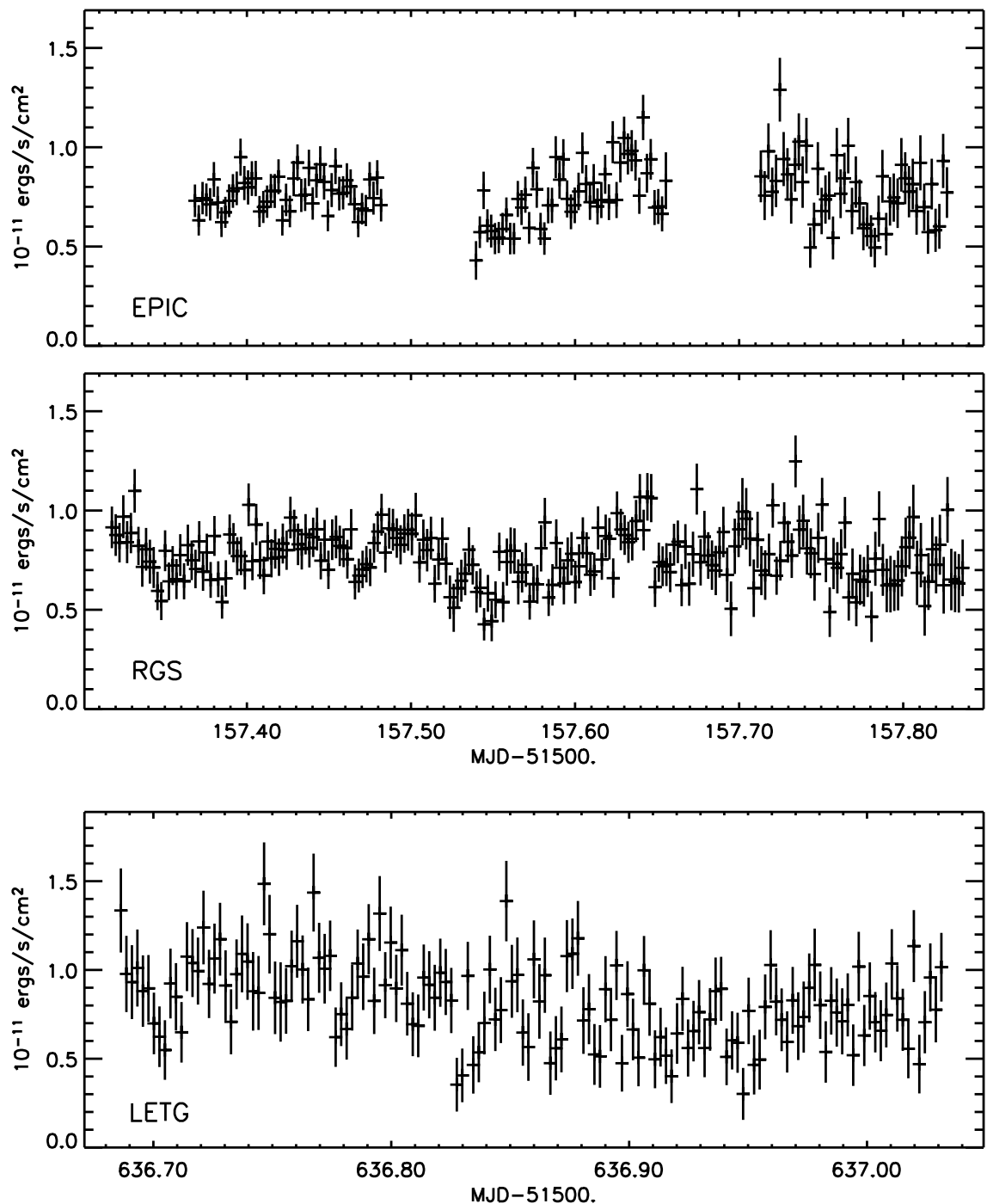}
\figurenum{2}
\caption{X-Ray light curves of CAL~83 built from XMM-\newton\  EPIC pn (top panel),
RGS (middle panel), and \chandra\  HRC-S/LETG1 observations. Note the short-timescale
variability, while the flux level remains the same between the XMM-\newton\  and the
\chandra\  observations.
\label{VarFig}}
\end{figure}

Figure~\ref{VarFig} shows the light curves of CAL~83 with XMM-\newton\  (EPIC pn: top panel,
RGS: middle panel) and \chandra\ (LETG1: bottom panel) in units of flux observed at Earth.
We use our final model spectrum in combination with the respective response matrices to obtain
conversion factors. Note that, in the case of EPIC pn, the observation was cut into three pieces 
of similar length but using different filters. To avoid pile-up and optical contamination, we
extracted EPIC events in an annulus ($0.2-0.8$~keV range), and used a nearby region for the background. 
In the case of RGS and LETG, the background was obtained from events ``above'' and ``below'' the 
dispersed spectrum. We used events in the range $20-37$~\AA\  for RGS, and $20-65$~\AA\  for LETG, 
since no signal was present outside these ranges.

Although the absolute observed flux still contains some uncertainty (e.g., cross-calibration between
instruments, imperfect model), similar flux levels were obtained during both the XMM-\newton\  
and \chandra\  observations, despite their time separation of $1.25$~yr. \cite{greiner02}
also noted that the flux observed by XMM-\newton\  was comparable to earlier observations with
ROSAT and {\em Beppo}SAX. Not only the flux levels were the same, but we found that the
RGS and the LETG spectra were remarkably similar (Fig.~\ref{LETGFig}). On the other hand,
we have detected relative, short-timescale variations in the X-ray light curves.
Their typical timescale is much shorter than the 1.04-day binary orbital period.
These flux variations can reach up to 50\% of the average flux. 
Hardness ratios light curves do not show significant variations,
suggesting that the flux variations are not due to temperature
fluctuations. However, because of the limited signal-to-noise ratio  due to the small effective
areas, we stress that we cannot exclude that temperature effects played some role in the flux
variations. Simultaneous UV and optical photometry with a similar time sampling would be
necessary to study stochastic effects in the accretion process. 


\section{NLTE Model Atmospheres}
\label{NLTE}

We have constructed a series of NLTE line-blanketed model
atmospheres of hot WD's with our model atmosphere program,
TLUSTY, version 201. Detailed emergent spectra are then calculated
with our spectrum synthesis code, SYNSPEC, version 48, using
the atmospheric structure and the NLTE populations previously
obtained with TLUSTY. TLUSTY computes stellar model photospheres
in a plane-parallel geometry, assuming radiative and hydrostatic
equilibria. Departures from LTE are explicitly allowed for
a large set of chemical species and arbitrarily complex model atoms,
using our hybrid Complete Linearization/Accelerated
Lambda Iteration method \citep{NLTE1}. This enables us to account
extensively for the line opacity from heavy elements, an essential
feature as the observed spectrum is suggestive of strong line
opacity (see Fig.~\ref{LETGFig}).

We have implemented in our two codes several specific upgrades for
computing very hot model atmospheres as well as detailed X-ray
spectra. Starting with version 200, TLUSTY is a universal
code designed to calculate the vertical structure of stellar atmospheres
and accretion disks. With this unification, we directly
benefit from the upgrades implemented to calculate very hot accretion disk
annuli. This concerns in particular a treatment of opacities from highly
ionized metals and Compton scattering \citep{AGN4}. 

Originally, \citet{AGN4} implemented a simple description of metal opacities
to explore the basic effect of these opacity sources on accretion disk spectra.
Adopting one-level model atoms, they treated all ions, from neutrals to fully
stripped atoms, of the most abundant chemical species: H, He, C, N, O, Ne, Mg,
Si, S, Ar, Ca, and Fe. Therefore, only photoionization from the ground state
of each ion is considered. Photoionization cross-sections, including Auger
inner-shell photoionization, were extracted from the X-ray photoionization
code XSTAR \citep{kallman00}. We made the simplifying assumption that, if an Auger
transition is energetically possible, then it occurs and the photoionization
results in a jump by two stages of ionization to a ground state configuration.
Fluorescence and multiple Auger electron ejection arising from inner shell photoionization
are neglected. To handle dielectronic recombination, we followed the description
of \citet{AGN4}, based on data from \citet{aldrovandi73}, \citet{nussbaumer83}, and
\citet{arnaud92}, which are used in XSTAR.

These models only include bound-free opacities from the ground states. Therefore,
we constructed multi-level model atoms in order to account for the effect of metal lines
in our NLTE model atmospheres. We did so only for the most populated ions, and ions
with very low populations are excluded. We retained
the ionization and recombination data discussed above for ground state levels, and
expanded the model atoms using data calculated by the Opacity Project \citep{IOP95, IOP97}.
It is essential to explicitly incorporate highly-excited levels in the model atoms,
because X-ray lines (where the model atmosphere flux is maximal) are transitions between
low-excitation (or ground state) and these highly-excited levels. To deal with the large
number of excited levels, we merged levels with close energies into superlevels
assuming that they follow Boltzmann statistics relative to each other (that is, all
individual levels in a superlevel share the same NLTE departure coefficient). Proper
summing of individual transitions is applied for transitions between superlevels.
Refer to \citet{NLTE1} and \citet{OS02} for details on NLTE superlevels and on the
treatment of transitions in TLUSTY.  Table~\ref{IonTbl} lists the ions included in
the model atmospheres, the number of explicit NLTE levels and superlevels, and the
corresponding numbers of superlevels and lines. We recall here that OP neglected the
atomic fine structure, and these figures refer to OP data (hence, the actual number
of levels and lines accounted for is higher by a factor of a few). For each ion, we
list the original publications unless the calculations were only published as a part
of the OP work. Data for one-level model atoms are from \cite{kallman00}.

The second upgrade in TLUSTY deals with electron scattering, and we have considered
the effect of Compton {\it vs.\/} Thompson scattering. At high energies, Compton scattering
provides a better physical description.  Compton scattering is incorporated in the radiative
transfer equation in the nonrelativistic diffusion approximation through a Kompaneets-like
term. Details of the implementation are described in \citet{AGN4}. 
Through a coupling in frequencies, this increases significantly the computational cost.
We thus decided to explore the differences resulting from using Compton or Thompson scattering.
We performed
a small number of tests comparing model atmospheres and predicted spectra which were computed
using these two approaches. At the considered
temperatures ($T\approx 500,000$\,K), the differences are very small and become visible
only in the high-energy tail, above 2\,keV. At these energies, the predicted
flux is very low and, indeed, no flux has been observed in CBSS at these energies.
Moreover, the changes result in little feedback on the calculated atmospheric structure.
We may therefore safely use Thompson scattering in modeling CBSS atmospheres.

Detailed spectra are produced in a subsequent step with our spectrum synthesis
program SYNSPEC, assuming the atmospheric structure and NLTE populations calculated
with TLUSTY. Upgrades made in the TLUSTY program have been transported in
SYNSPEC when necessary. A detailed line list for the soft X-ray domain was built,
combining essentially two sources. The initial list was extracted from Peter van Hoof
Atomic Line List\footnote{http://www.pa.uky.edu/$\sim$peter/atomic/}. This list contains
transitions between levels with measured energies\footnote{from the NIST Atomic Spectra
Database at
http://physics.nist.gov/cgi-bin/AtData/main\_asd}, and all lines thus have accurate
wavelengths. These lines represent, however, a small fraction of all lines in the
soft X-ray range because the energy of most highly-excited levels has not been
measured from laboratory spectra. We thus complemented this initial list with a complete
list of transitions from OP. The OP lines do not account for fine structure, and wavelengths
are derived from theoretical energies. The expected accuracy of the theoretical 
wavelengths is expected to be of the order of 0.5\,\AA\  or better. This is, however,
not as good as the spectral resolution
achieved with {\sl Chandra\/} or XMM-{\sl Newton\/} spectrometers, and potential difficulties may
be expected when comparing model spectra to observations. Yet the\ global opacity effect
of these lines is important. The line list only contains lines
from the explicit NLTE ions. The spectra are calculated using the
NLTE populations from the TLUSTY models, and SYNSPEC does not require partition
functions in this case.
Finally, we examined the issue of line broadening. As discussed by \citet{SA78},
natural broadening becomes the main effect for lines of highly-charged ions, and
already dominates the linear Stark broadening for \ion{C}{6} lines. Therefore,
we included natural broadening data when available, and we neglected Stark broadening.

\begin{figure}
\epsscale{.84}
\plotone{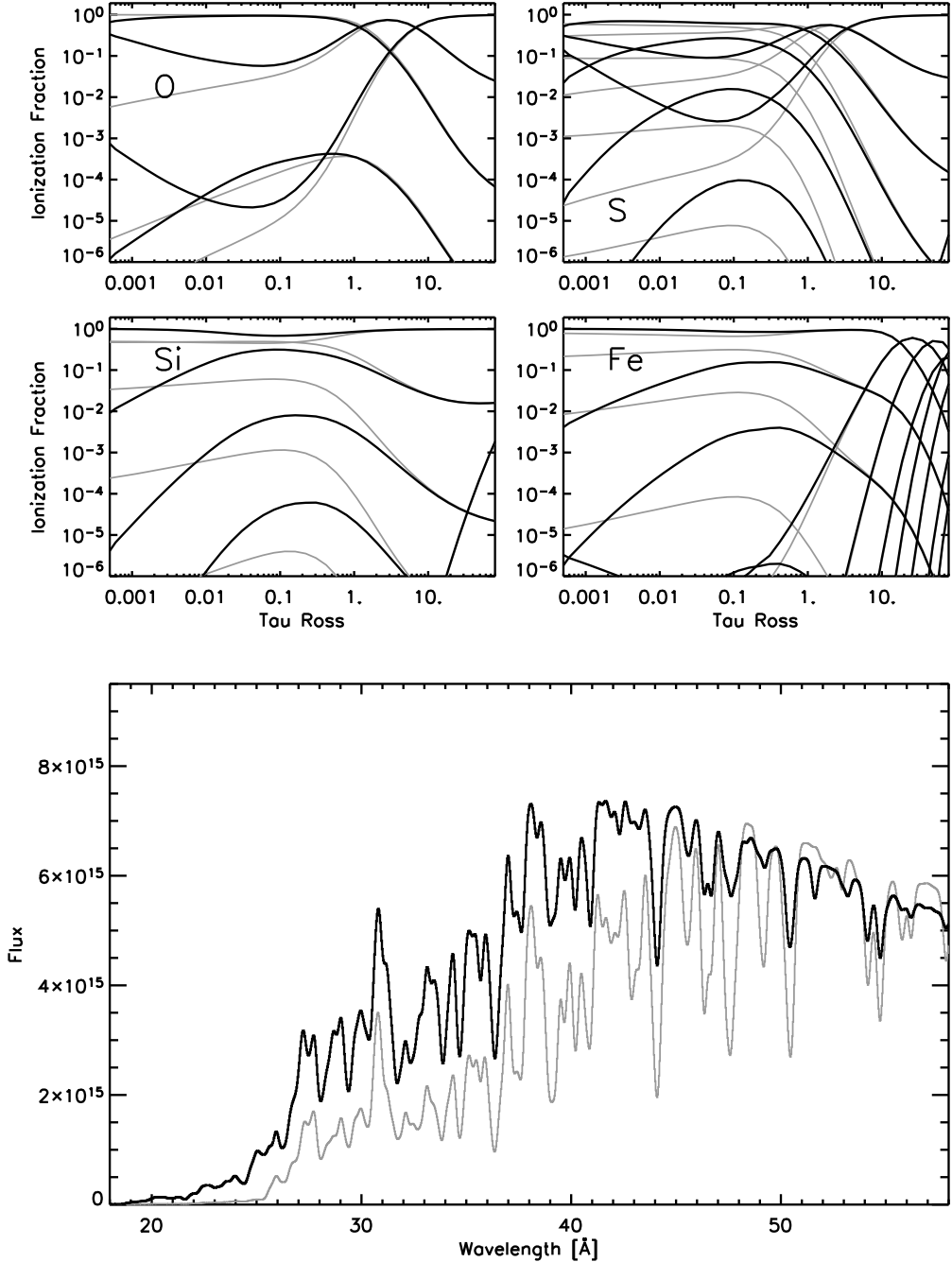}
\figurenum{3}
\caption{Ionization fractions and predicted spectra of a NLTE (black lines) and
a LTE (grey lines) model atmosphere, \teff = 550,000\,K, $\log g = 8.5$. The top
part shows the ionization fractions of O, Si, S and Fe. Dominant ions
in the line formation region ($\tau_{\rm Ross}\approx 0.1 - 0.01$) are \ion{O}{7},
\ion{Si}{13}, \ion{S}{13} and \ion{Fe}{17}. LTE and NLTE predicted spectra,
convolved to a 0.3\,\AA\  resolution, are compared in the bottom panel.  
\label{LTEFig}}
\end{figure}

Departures from LTE are illustrated in Fig.~\ref{LTEFig} for a model atmosphere with
\teff\  = 550,000\,K, $\log g = 8.5$, and a LMC metallicity. The ionization fractions
show that the NLTE model is overionized compared to the LTE model, which is a typical behavior
of NLTE model stellar atmospheres where the radiation temperature is higher than the
local electronic temperature. As expected, the NLTE ionization fractions return to their
LTE values at depth ($\tau_{\rm Ross}\ga 2$). The overionization remains limited because
the local temperatures are higher in the LTE model compared to the NLTE model as a result
of stronger continuum and line absorption at depths around $10^{-3}\la\tau_{\rm Ross}\la 1$.
This stronger absorption is indeed visible in the predicted emergent spectrum: the
LTE predicted flux is significantly lower below 45\,\AA. Deriving a lower surface gravity,
hence a lower WD mass, would therefore be a consequence of a LTE analysis compared to the
NLTE analysis (see also \S\ref{Resu} and Fig.~\ref{MFitFig}).


\section{Spectrum Analysis}
\label{Analysis}

\subsection{Methodology}
\label{Method}

\begin{figure}
\epsscale{.99}
\plotone{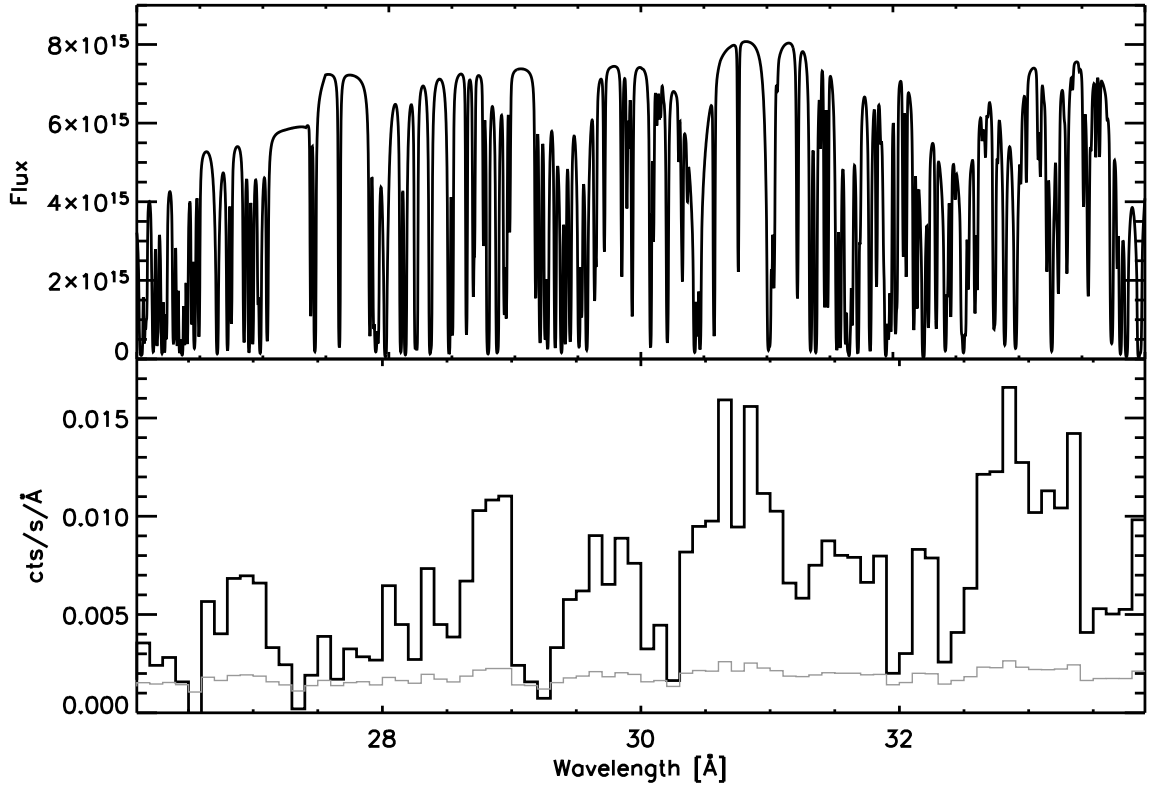}
\figurenum{4}
\caption{Comparison of the observed spectral resolution in the \chandra\  LETG
spectrum (bottom panel; observed counts~s$^{-1}$~\AA$^{-1}$) with the best fit
model spectrum (top panel; flux at the stellar surface in ergs~s$^{-1}$~cm$^{-2}$~\AA$^{-1}$),
demonstrating that the observed spectral features are blends of many metal lines.
The 1-$\sigma$ count rate error is shown in the bottom panel (thin grey line) allowing
to assess whether the observed spectral features are genuine or not.
\label{ResolFig}}
\end{figure}

We display in Fig.~\ref{ResolFig} a typical spectral region of the \chandra\  LETG data
together with the predicted spectrum of the stellar surface, that is without any instrumental
convolution. This clearly illustrates that all observed features are actually blends of
many metal lines, mostly from \ion{Si}{12}, \ion{S}{11}, \ion{S}{12}, \ion{Ar}{10},
\ion{Ar}{11}, \ion{Ar}{12}, \ion{Ca}{12}, and \ion{Fe}{16}.
It is thus impractical to identify these features individually. Moreover,
the fact that we had to use theoretical wavelengths for many predicted lines compounds the
difficulties. Therefore, the spectrum analysis cannot be based on fitting a few key features
to derive the properties of CAL~83, but we rather need to achieve an overall match to the
observed spectrum. To this end, we do not use any statistical criterion but we simply match
the model to the data by eye. A $\chi^2$ statistics would primarily measure the line opacity
that is missing in the model (a systematic effect) rather than efficiently discriminate between
different values of the stellar parameters (see \S\ref{Resu}). Errors are analogously estimated
by eye, that is, models outside the error box clearly exhibit a poor match to the observations.
We feel that this approach provides conservative error estimates.

We started from the stellar parameters derived by \cite{paerels01}, and built a small
grid of NLTE model atmospheres around these initial estimates covering a range in effective
temperature and surface gravity, $450,000\la T_{\rm eff}\la 600,00$\,K and
$8\la \log g\la 9$. For each set of parameters, we constructed the final model in
a series of models of increasing sophistication, adding more explicit NLTE species, ions,
levels, and lines. Table~\ref{IonTbl} lists the explicit species included in the final models.
This process was helpful numerically to converge the models, but this was primarily used
to investigate the resulting effect on the predicted spectrum of incorporating additional ions
in the model atmospheres. We have assumed that the surface composition of CAL~83 reflects
the composition of the accreted material, i.e. typical of the LMC composition. We have
adopted a LMC metallicity of half the solar value \citep{rolleston02}. Additional models with
a composition reflecting an evolved star donor with CNO-cycle processed material and
enhanced $\alpha$-element abundances have also been calculated to explore the effect of
changing the surface composition on the derived stellar parameters.

To compare the model spectra to the \chandra\  LETG and XMM-\newton\  RGS data,
we applied to the model a correction for the interstellar (IS)
extinction, the appropriate instrumental response matrix (see \S3),
and a normalization factor $(R_{\rm WD}/d)^2$.

Our IS model is based on absorption cross-sections in the X-ray domain compiled
by \cite{BCMC92} for 17 astrophysically important species. Specifically,
the effective extinction curve was calculated with Balucinska-Church~\&
McCammon's code\footnote{Available at http://cdsweb.u-strasbg.fr/viz-bin/Cat?VI/62},
assuming a solar abundance mix \citep{grevesse98}. The total extinction is
then proportional to the hydrogen column density. We used the value,
$N_{\rm H} = 6.5\pm 1.0~10^{20}$~cm$^{-2}$, measured by \cite{gansicke98}
from Ly\,$\alpha$ {\sl HST\/} GHRS observations of CAL~83.

After the IS extinction and instrumental corrections, the model spectra were
scaled to the \chandra\  LETG spectrum to match the observed flux level longward of
45\,\AA, yielding the normalization factor $(R_{\rm WD}/d)^2$. The WD radius
immediately follows from the adopted distance to the LMC, $d = 50\pm 3$~kpc,
based on RR~Lyrae and eclipsing binaries \citep{alcock04, clausen03}.


\subsection{Results and Discussion}
\label{Resu}

\begin{figure}
\epsscale{.99}
\plotone{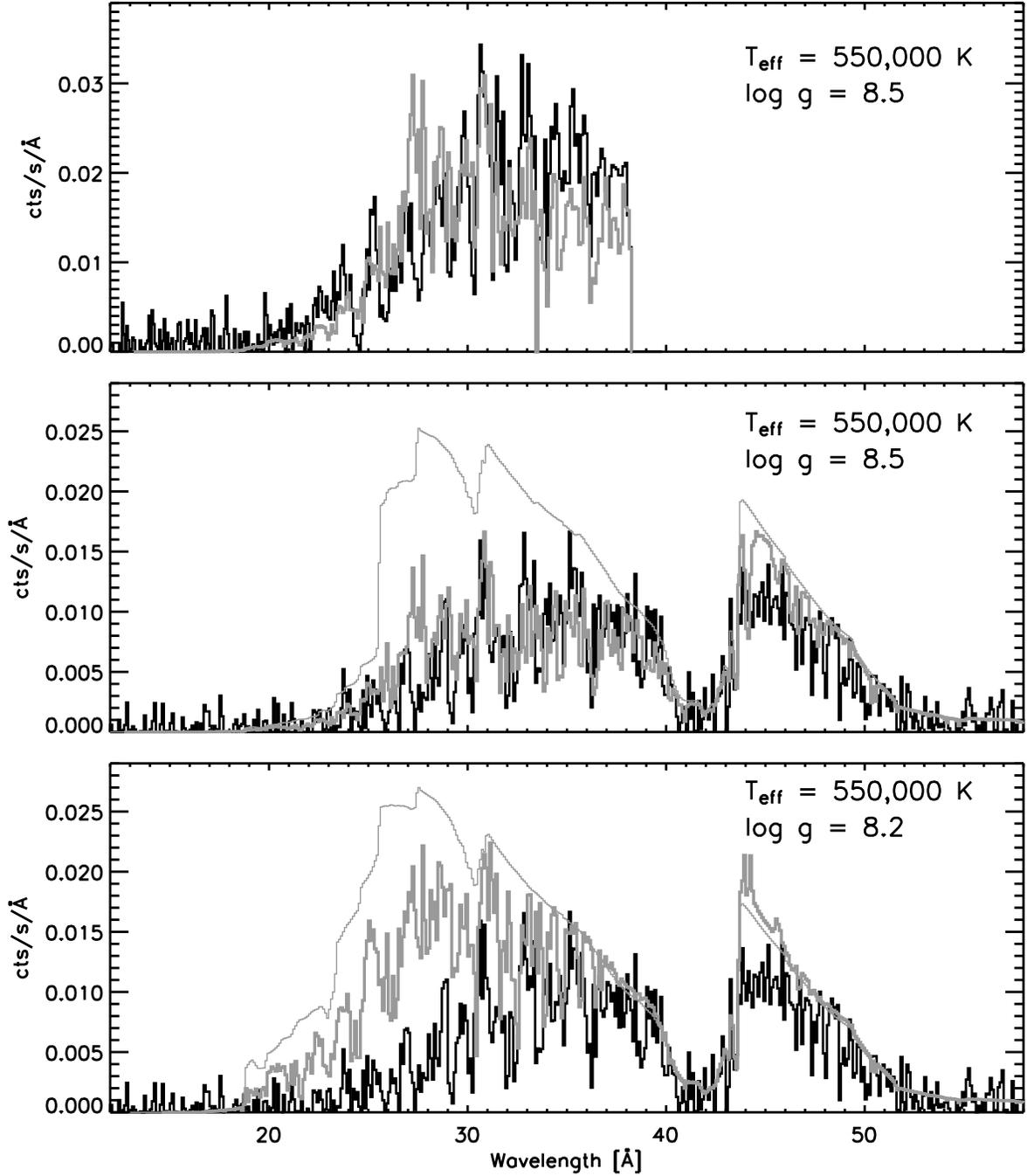}
\figurenum{5}
\caption{Best model fit to the RGS (top) and LETG (middle) spectra. The model spectrum
is normalized to the LETG spectrum between 45 and 50\,\AA, and the same normalization
is applied when matching the RGS spectrum. This normalization yields a WD radius of
0.01\,$R_\odot$ and a WD mass of 1.3\,$M_\odot$ (see Table~\ref{ResuTbl}). The bottom
panel shows the model sensitivity to surface gravity, hence to the WD mass, that is here
0.65\,$M_\odot$. The predicted continuum spectra are also shown to
illustrate the importance of line opacity.  \label{MFitFig}}
\end{figure}

\begin{figure}
\epsscale{.99}
\plotone{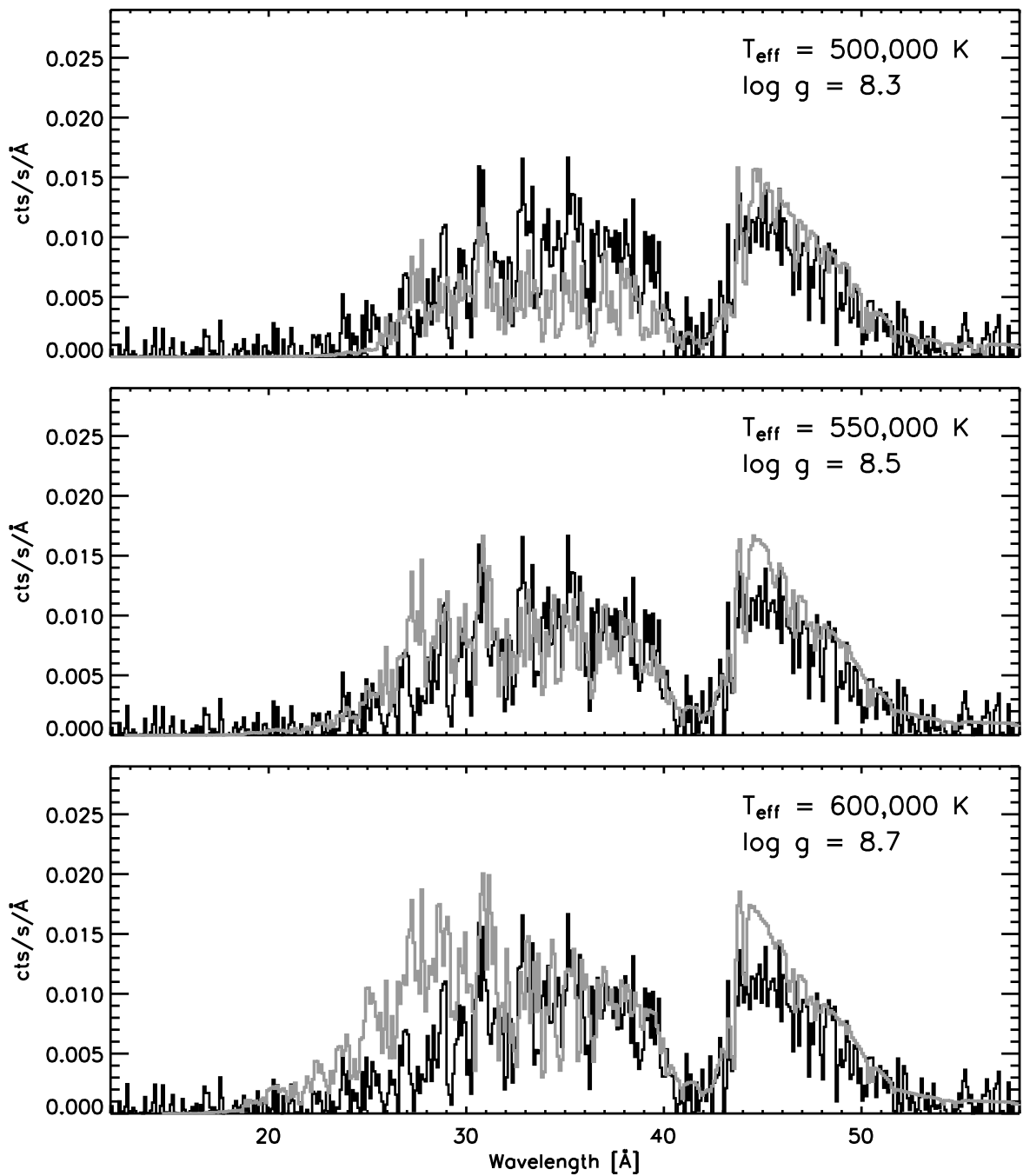}
\figurenum{6}
\caption{Comparison of the best model fit (middle panel) to the LETG spectrum with
cooler (top) and hotter (bottom) models. The model spectra are normalized to the LETG
spectrum between 45 and 50\,\AA, and the derived WD radius decreases with increasing \teff.
All three models have a WD mass of about 1.3\,$M_\odot$. 
\label{TFitFig}}
\end{figure}

We present in Fig.~\ref{MFitFig} and~\ref{TFitFig} our best model fit to the RGS and LETG
spectra along with the model sensitivity to \teff\  and $\log g$. The best model spectrum
has been normalized to match the observed LETG flux between 45 and 50\,\AA, yielding a WD
radius of $R_{\rm WD} = 7.3~10^8$\,cm~$\approx 0.01\,R_\odot$. The same normalization factor
is then applied for comparing the model spectrum to the RGS spectrum. The benefit
of having a broader spectral coverage with LETG is instantly apparent for deriving
an accurate model normalization. The top two panels
of Fig.~\ref{MFitFig} generally show a satisfactory agreement between the model and
the two observed spectra. Above 34\,\AA, the match to the RGS spectrum is not as good as
for the LETG spectrum, probably resulting from the difficulty of correcting the RGS data
for fixed-pattern noise in this range. The general energy distribution and most spectral
features are well reproduced, for instance the features at 24, 29, 30, 32, and 36\,\AA.
On the other hand, the observed absorption at $\approx$~27\,\AA\  is not reproduced by our model.
A cursory look might suggest that the model predicts an emission feature there but,
overplotting the predicted continuum flux, we see that the model essentially misses line opacity
around 27-28\,\AA. In this respect, our analysis definitively demonstrates that we observe
a photospheric absorption spectrum, with no obvious evidence of emission lines. 

Fig.~\ref{MFitFig} presents two models with a different surface gravity, clearly demonstrating
that our data allow us to determine $\log g$. The higher flux in the low gravity model is
the result of a higher ionization in the low gravity model atmosphere, decreasing the
total opacity in the range between 20 and 30\,\AA. This effect is large enough to determine
the surface gravity with a good accuracy, typically $\pm~0.1$~dex on $\log g$.
Fig.~\ref{TFitFig} shows three models with different effective temperatures. The surface
gravity was adjusted to provide the closest match to the observed spectrum, and the
appropriate model normalizations and WD radii were determined. The differences between
the models are not as large as in the case of gravity, but we may
exclude models as cool as 500,000\,K or as hot as 600,000\,K. We thus adopt
\teff~=~550,000~$\pm$~25,000\,K ($k$\teff~=~46~$\pm$~2~eV). The WD mass and luminosity
straightforwardly follow, and our results are summarized in Table~\ref{ResuTbl}.
Uncertainties are propagated as dependent errors, formally yielding a WD mass of
$M_{\rm WD} = 1.3\pm 0.3 M_\odot$. The mass-radius relation for cold WD's predicts
a WD radius of about 0.004~$R_\odot$ for such a high WD mass \citep{hamada61}, implying
that CAL~83 has a swollen atmosphere with a measured radius 2.5 times the expected value.
\citet{kato97} calculations of WD envelopes predict an even larger radius ($\approx 0.025 R_\odot$)
for the derived mass and temperature, suggesting that the observed emission may emerge 
only from a hot cap on the WD surface. The derived bolometric luminosity,
$L_{\rm bol} = 3.5\pm1.2~10^{37}$~ergs~s$^{-1}$, is higher than the luminosity range given
by \cite{gansicke98} but consistent with \cite{kahabka98} estimate, and corresponds
to about 0.3~$L_{\rm Edd}$.

The surface composition of WD's in CBSS is yet unknown. It might reflect either the composition
of the accreted material or the nucleosynthetic yields of nuclear burning close to the WD
surface. Characterizing this composition would be of major interest for understanding
the physical processes occurring on the accreting WD. However, our first concern is to
verify how sensitive the determination of stellar parameters is with respect to the
assumed composition. So far, we have assumed a composition typical of the LMC metallicity.
If the donor star has already evolved off the main sequence, its surface composition might
already be altered and might show enhanced abundances of $\alpha$~elements and/or CNO
abundances resulting from the mixing with CNO-cycle processed material. We have therefore
constructed a NLTE model atmosphere, assuming the same effective temperature
and surface gravity, but all $\alpha$~elements are overabundant by a factor of 2, carbon
(1/50 the LMC abundance) and oxygen (1/5) are depleted, and nitrogen (8 times) is enhanced.
Fig.~\ref{ChemFig} compares the match of the two models to the LETG spectrum. A careful
examination shows that indeed the model with the altered composition reveals some stronger
absorption lines, but the changes remain small. We conclude first that the determination of
the stellar parameters is little affected by the assumed surface composition (within some
reasonable range; exotic compositions will not match the observations), and second that
a detailed chemical composition analysis would require improved atomic data.

Velocities are a further factor affecting the strength of the absorption line spectrum.
In particular, we adopted a microturbulent velocity of 50~km\,s$^{-1}$. The ratio of this
value to the sound speed is rather typical of values derived in other hot stars.
Nevertheless, we varied the microturbulent velocity from low values up to values close
to the sound velocity to investigate the effect of different microturbulences on the predicted
spectrum. The line absorption spectrum indeed strengthened at larger microturbulent velocities,
but the overall match to the LETG data is not much changed. The derived parameters remain,
therefore, insensitive to the assumed microturbulent velocity.
We also explored the role of an outflow in strengthening and broadening the spectral lines
because of the presence of a strong wind from the WD has been proposed to regulate the
accretion \citep{hachisu96}. We carefully examined the LETG spectrum, using different binning
factors. We did not find evidence of emission or line asymmetries that could reveal an outflow.
Note, however, that the spectral resolution would only allow us to detect
fast outflows ($>$2000~km\,s$^{-1}$). For the same reason, we can only exclude very fast
{\it apparent} rotation velocities, e.g., values close to the critical velocity
($\approx$\,7000~km\,s$^{-1}$). It is likely, however, that the inclination angle of the WD
rotation axis is small because the absence of eclipses indicates that the accretion disk
is almost seen face-on. Therefore, our observations cannot rule out a fast WD rotation which
is advocated by \citet{yoon04} for the WD to gain mass efficiently.

The uncertainty on the IS transmission might be the most severe issue. Our IS
model (see \S\ref{Method}) predicts a transmission as low as 6\% at 50\,\AA\ increasing
up to about 60\% at 20\,\AA. A change in the assumed hydrogen column density toward
CAL~83 might thus significantly affect the results. Fortunately, this column was measured
accurately from Ly\,$\alpha$ observations by \cite{gansicke98},
$N_{\rm H} = 6.5\pm 1.0~10^{20}$~cm$^{-2}$. We use their
error range to explore the resulting changes. Thanks to the relatively tight range, the
spectral shape between 25 and 40\,\AA\  is little changed, and stays within changes predicted
from the adopted errors on \teff\   and $\log g$. Therefore, the error on the column
density primarily translates into different normalization factors, i.e. a range of
WD radii. A higher column density
yields a larger WD radius, hence a larger WD mass (which should not exceed \Mch).
Conversely, a smaller column density yields a lower WD radius, hence a lower luminosity.
Since the measured luminosity is close the boundary of stable steady-state nuclear
burning \citep{iben82, vdH92}, it is unlikely that the WD luminosity is much smaller.
Therefore, the actual value of the IS density column likely is within an even tighter range.
The uncertainty on the WD radius listed in Table~\ref{ResuTbl} remains relatively small, because
it only combines the uncertainty on the normalization factor and on the LMC distance which
are well determined. Fully accounting for \cite{gansicke98} error bar on $N_{\rm H}$ would
translate to unrealistically large errors on the WD radius and on the WD mass.

We revisit now the issue of the X-ray off-states in the light of our result that the WD
has a high mass. \citet{kato97} calculations of WD envelopes showed that the X-ray turnoff
time after the cessation of nuclear burning could be very short for massive WDs. An observed
turnoff time as short as 20 days \citep{kahabka98} implies $M_{\rm WD} \ga 1.35 M_\odot$.
\cite{gansicke98} noted that CAL~83 has a low luminosity for a CBSS and is found close to
the stability limit of steady burning, which is thus consistent with the idea of unstable
burning and might also be related to our discovery of short term variability of the X-ray flux.
Although the actual process responsible of the off-states cannot be definitively
established, the characteristic timescale supports the idea of a massive WD in CAL~83.


\section{Conclusions}
\label{Conclu}

Table~\ref{ResuTbl} summarizes our results. Our spectroscopic analysis indicates that CAL~83
contains a massive WD, $M_{\rm WD} = 1.3\pm 0.3 M_\odot$. Obviously, the upper limit is
well over \Mch, but a more interesting issue to tie CBSS and SN~Ia progenitors is the
lower limit for the mass.  Because low-mass models do not provide a match to the observed
spectrum that is as good as the fits achieved with high-mass models, we have concluded that
$M_{\rm WD} > 1.0~M_\odot$ is a robust lower limit for the WD mass. Nevertheless,
we need to emphasize here that this conclusion depends on our current description of the
soft X-ray opacities, particularly between 20 and 40\,\AA. We believe that the OP data
already provide a good description and, although progress in this regard
would be very helpful, we do not expect changes drastic enough to modify our conclusion.
The issue of IS extinction is potentially more serious, but we have argued that
we have an accurate estimate of the hydrogen column density toward CAL~83. In addition,
our model is consistent with the WD surface having a hydrogen-rich composition with LMC
metallicity. Finally, our analysis does not reveal any evidence of an outflow from the
WD.

We have reported a third X-ray off-state, showing that CAL~83 is a source undergoing unstable
nuclear burning. This is consistent with its low luminosity. The short timescale of
the off-states (about 50 days) provides a supporting evidence that CAL~83 WD is massive
($M_{\rm WD}\ga 1.35 M_\odot$). A better characterization of the off-states is required
to definitively establish the mechanism(s) responsible of the off-states.
Moreover, we confirm that the X-ray luminosity is fairly constant on a long-term basis
during on-states. However, on the other hand, we found variations up to 50\% of the soft
X-ray luminosity in very short times (less than 0.1 orbital period). Simultaneous X-ray and
optical photometry covering continuously a duration longer than a full orbital period would
be very valuable to study stochastic accretion processes.

Within the model of SN~Ia progenitors proposed by \cite{hachisu96}, our results would 
place CAL~83 in a late stage, after the strong accretion and wind phase when the accretion rate
drops below the critical rate for sustaining steady nuclear burning. Our results therefore
make CAL~83 a very likely candidate for a future SN~Ia event. We plan to conduct a similar
NLTE analysis of the few other CBSS already observed with \chandra\  LETG spectrometer to
further support our results and quantitatively study differences between these objects.
In particular, the second typical CBSS, CAL~87, exhibits a very different spectrum
characterized by emission lines \citep{greiner04}. A detailed spectroscopic analysis should
reveal either if the different spectrum arises from a different geometry (edge-on {\it vs.\/}
face-on system) or if CAL~87 is in an earlier evolutionary stage.

\acknowledgments

Tim Kallman kindly supplied photoionization cross sections implemented in
his code XSTAR. M.~A. is thankful to Kevin Briggs for insightful
information on non-detection upper limits. This work was supported by a grant
from the NASA Astrophysics Theory Program (NRA 00-01-ATP-153).
The Columbia group acknowledges support from NASA to Columbia University for 
XMM-\newton\  mission support and data analysis. T.~L. enjoyed the hospitality of
Columbia Astrophysics Laboratory during the completion of this work.


\clearpage

\clearpage

\clearpage

\clearpage

\clearpage

\clearpage

\clearpage

\clearpage

\begin{deluxetable}{llrlll}
\tablewidth{0pt}
\tabletypesize{\small}
\tablecaption{\chandra\  and XMM-\newton\  Observation Log of CAL 83. \label{LogTbl}}
\tablehead{  \colhead{Mission } & \colhead{Instrument } & \colhead{Exp. Time} &
             \colhead{Start Time} & \colhead{End Time}  & State }
\startdata
\chandra\     &  HRC-S/LETG &  52.3 ksec  &  1999-11-29~06:33 & 1999-11-29~21:27 & Off-state  \\*
\chandra\     &  ACIS-S     &   2.1 ksec  &  1999-11-30~14:50 & 1999-11-30~15:47 & Off-state  \\*
XMM-\newton\  &  RGS \& EPIC&  45.1 ksec  &  2000-04-23~07:34 & 2000-04-23~20:04 & \\*
\chandra\     &  HRC-S/LETG &  35.4 ksec  &  2001-08-15~16:03 & 2001-08-16~02:10 &  \\*
\chandra\     &  HRC-S/LETG &  61.6 ksec  &  2001-10-03~11:35 & 2001-10-04~05:12 & Off-state  \\*
\enddata
\end{deluxetable}

\singlespace
\clearpage

\begin{deluxetable}{lrrrl}
\tabletypesize{\small}
\tablecaption{Atomic data included in the NLTE model atmospheres. \label{IonTbl}}
\tablehead{
\colhead{Ion} & \colhead{(Super)Levels}   & \colhead{Indiv. Levels}   &
\colhead{Lines} & \colhead{References}
}
\startdata
\ion{H}{1}  &   9\phm{Levels}  &  80\phm{Levels}  &     172\phm{a}  &                \\*
\ion{H}{2}  &   1\phm{Levels}  &   1\phm{Levels}  &                 &                \\ [2mm]
\ion{He}{2} &  15\phm{Levels}  &  15\phm{Levels}  &      105\phm{a}  &                \\*
\ion{He}{3} &   1\phm{Levels}  &   1\phm{Levels}  &                 &                \\ [2mm]
\ion{C}{5}  &   9\phm{Levels}  &  19\phm{Levels}  &      27\phm{a}  & \phm{aaa}\citet{ADOC7}  \\*
\ion{C}{6}  &  15\phm{Levels}  &  15\phm{Levels}  &     105\phm{a}  &                \\*
\ion{C}{7}  &   1\phm{Levels}  &   1\phm{Levels}  &                 &                \\ [2mm]
\ion{N}{5}  &   5\phm{Levels}  &  14\phm{Levels}  &      28\phm{a}  & \phm{aaa}\citet{ADOC9}  \\*
\ion{N}{6}  &   9\phm{Levels}  &  19\phm{Levels}  &      27\phm{a}  & \phm{aaa}\citet{ADOC7}  \\*
\ion{N}{7}  &  15\phm{Levels}  &  15\phm{Levels}  &     105\phm{a}  &                \\*
\ion{N}{8}  &   1\phm{Levels}  &   1\phm{Levels}  &                 &                \\ [2mm]
\ion{O}{5}  &   1\phm{Levels}  &   1\phm{Levels}  &                 &                \\*
\ion{O}{6}  &   8\phm{Levels}  &  20\phm{Levels}  &      60\phm{a}  & \phm{aaa}\citet{ADOC9}  \\*
\ion{O}{7}  &  13\phm{Levels}  &  41\phm{Levels}  &     107\phm{a}  & \phm{aaa}\citet{ADOC7}  \\*
\ion{O}{8}  &  15\phm{Levels}  &  15\phm{Levels}  &     105\phm{a}  &                \\*
\ion{O}{9}  &   1\phm{Levels}  &   1\phm{Levels}  &                 &                \\ [2mm]
\ion{Ne}{5} &   1\phm{Levels}  &   1\phm{Levels}  &                 &                \\*
\ion{Ne}{6} &   1\phm{Levels}  &   1\phm{Levels}  &                 &                \\*
\ion{Ne}{7} &   1\phm{Levels}  &   1\phm{Levels}  &                 &                \\*
\ion{Ne}{8} &   5\phm{Levels}  &  14\phm{Levels}  &      28\phm{a}  & \phm{aaa}\citet{ADOC9}  \\*
\ion{Ne}{9} &   9\phm{Levels}  &  19\phm{Levels}  &      27\phm{a}  & \phm{aaa}\citet{ADOC7}  \\*
\ion{Ne}{10} &  15\phm{Levels}  &  15\phm{Levels}  &     105\phm{a}  &                \\*
\ion{Ne}{11} &   1\phm{Levels}  &   1\phm{Levels}  &                 &               \\ [2mm]
\ion{Mg}{9} &   1\phm{Levels}  &   1\phm{Levels}  &                 &                \\*
\ion{Mg}{10} &  12\phm{Levels}  &  54\phm{Levels}  &     284\phm{a}  & \phm{aaa}\citet{ADOC9}  \\*
\ion{Mg}{11} &  15\phm{Levels}  & 109\phm{Levels}  &     565\phm{a}  & \phm{aaa}\citet{ADOC7}  \\*
\ion{Mg}{12} &   1\phm{Levels}  &   1\phm{Levels}  &                 &               \\ [2mm]
\ion{Si}{9} &   22\phm{Levels}  & 489\phm{Levels}  &   11900\phm{a}  & \phm{aaa}\citet{ADOC11}  \\*
\ion{Si}{10} &  24\phm{Levels}  & 313\phm{Levels}  &    7531\phm{a}  & \phm{aaa}\citet{IOP95}  \\*
\ion{Si}{11} &  16\phm{Levels}  & 184\phm{Levels}  &    2599\phm{a}  & \phm{aaa}\citet{ADOC14}  \\*
\ion{Si}{12} &  12\phm{Levels}  &  54\phm{Levels}  &     284\phm{a}  & \phm{aaa}\citet{ADOC9}  \\*
\ion{Si}{13} &   9\phm{Levels}  &  19\phm{Levels}  &      27\phm{a}  & \phm{aaa}\citet{ADOC7}  \\*
\ion{Si}{14} &   1\phm{Levels}  &   1\phm{Levels}  &                 &               \\ [2mm]
\ion{S}{9} &   20\phm{Levels}  &  481\phm{Levels}  &   13665\phm{a}  & \phm{aaa}\citet{IOP95}  \\*
\ion{S}{10} &  23\phm{Levels}  &  512\phm{Levels}  &   17966\phm{a}  &  \phm{aaa}\citet{IOP95}  \\*
\ion{S}{11} &  23\phm{Levels}  &  587\phm{Levels}  &   16253\phm{a}  &  \phm{aaa}\citet{ADOC11}  \\*
\ion{S}{12} &  23\phm{Levels}  &  301\phm{Levels}  &    7181\phm{a}  &  \phm{aaa}\citet{IOP95}  \\*
\ion{S}{13} &  13\phm{Levels}  &  184\phm{Levels}  &    2599\phm{a}  & \phm{aaa}\citet{ADOC14}  \\*
\ion{S}{14} &  12\phm{Levels}  &   54\phm{Levels}  &     284\phm{a}  & \phm{aaa}\citet{ADOC9}  \\*
\ion{S}{15} &   9\phm{Levels}  &   19\phm{Levels}  &      27\phm{a}  & \phm{aaa}\citet{ADOC7}  \\*
\ion{S}{16} &   1\phm{Levels}  &   1\phm{Levels}  &                 &               \\ [2mm]
\ion{Ar}{8} &   1\phm{Levels}  &    1\phm{Levels}  &                 &                \\*
\ion{Ar}{9} &   23\phm{Levels}  & 197\phm{Levels}  &    2877\phm{a}  &  \phm{aaa}\citet{IOP95}  \\*
\ion{Ar}{10} &  19\phm{Levels}  & 287\phm{Levels}  &    6836\phm{a}  &  \phm{aaa}\citet{IOP95}  \\*
\ion{Ar}{11} &  20\phm{Levels}  & 514\phm{Levels}  &   15943\phm{a}  &  \phm{aaa}\citet{IOP95}  \\*
\ion{Ar}{12} &  22\phm{Levels}  & 546\phm{Levels}  &   21675\phm{a}  &  \phm{aaa}\citet{IOP95}  \\*
\ion{Ar}{13} &  26\phm{Levels}  & 617\phm{Levels}  &   19248\phm{a}  &  \phm{aaa}\citet{ADOC11}  \\*
\ion{Ar}{14} &  19\phm{Levels}  & 355\phm{Levels}  &    9755\phm{a}  &  \phm{aaa}\citet{IOP95}  \\*
\ion{Ar}{15} &  16\phm{Levels}  & 184\phm{Levels}  &    2599\phm{a}  &  \phm{aaa}\citet{ADOC14}  \\*
\ion{Ar}{16} &   1\phm{Levels}  &   1\phm{Levels}  &                 &               \\ [2mm]
\ion{Ca}{9} &   1\phm{Levels}  &   1\phm{Levels}  &                 &                \\*
\ion{Ca}{10} &  1\phm{Levels}  &   1\phm{Levels}  &                 &               \\*
\ion{Ca}{11} &  24\phm{Levels}  & 203\phm{Levels}  &    3098\phm{a}  &  \phm{aaa}\citet{IOP95}  \\*
\ion{Ca}{12} &  15\phm{Levels}  & 308\phm{Levels}  &    8043\phm{a}  &  \phm{aaa}\citet{IOP95}  \\*
\ion{Ca}{13} &  26\phm{Levels}  & 544\phm{Levels}  &   17703\phm{a}  &  \phm{aaa}\citet{IOP95}  \\*
\ion{Ca}{14} &  22\phm{Levels}  & 629\phm{Levels}  &   27835\phm{a}  &  \phm{aaa}\citet{IOP95}  \\*
\ion{Ca}{15} &  24\phm{Levels}  & 730\phm{Levels}  &   24897\phm{a}  &  \phm{aaa}\citet{ADOC11}  \\*
\ion{Ca}{16} &  21\phm{Levels}  & 381\phm{Levels}  &   11511\phm{a}  &  \phm{aaa}\citet{IOP95}  \\*
\ion{Ca}{17} &   1\phm{Levels}  &   1\phm{Levels}  &                 &               \\*
\ion{Ca}{18} &   1\phm{Levels}  &   1\phm{Levels}  &                 &               \\*
\ion{Ca}{19} &   1\phm{Levels}  &   1\phm{Levels}  &                 &               \\ [2mm]
\ion{Fe}{13} &   1\phm{Levels}  &   1\phm{Levels}  &                 &               \\*
\ion{Fe}{14} &   1\phm{Levels}  &   1\phm{Levels}  &                 &               \\*
\ion{Fe}{15} &  26\phm{Levels}  & 253\phm{Levels}  &    4797\phm{a}  &  \phm{aaa}\citet{ADOC19}  \\*
\ion{Fe}{16} &  17\phm{Levels}  &  52\phm{Levels}  &     285\phm{a}  &  \phm{aaa}\citet{IOP95}  \\*
\ion{Fe}{17} &  30\phm{Levels}  & 211\phm{Levels}  &    3409\phm{a}  &  \phm{aaa}\citet{IOP95}  \\*
\ion{Fe}{18} &   1\phm{Levels}  &   1\phm{Levels}  &                 &               \\
\ion{Fe}{19} &   1\phm{Levels}  &   1\phm{Levels}  &                 &               \\*
\ion{Fe}{20} &   1\phm{Levels}  &   1\phm{Levels}  &                 &               \\*
\ion{Fe}{21} &   1\phm{Levels}  &   1\phm{Levels}  &                 &               \\*
\ion{Fe}{22} &   1\phm{Levels}  &   1\phm{Levels}  &                 &               \\*
\ion{Fe}{23} &   1\phm{Levels}  &   1\phm{Levels}  &                 &               \\*
\ion{Fe}{24} &   1\phm{Levels}  &   1\phm{Levels}  &                 &               \\*
\ion{Fe}{25} &   1\phm{Levels}  &   1\phm{Levels}  &                 &               \\ [2mm]
\enddata
\end{deluxetable}
\clearpage

\begin{deluxetable}{ll}
\tablewidth{0pt}
\tablecaption{Stellar Parameters of CAL 83. \label{ResuTbl}}
\tablehead{  \colhead{ } & \colhead{ }   }
\startdata
Effective Temperature\phm{aaa}  &   $T_{\rm eff} = 5.5\pm 0.25~10^5$~K          \\*
Surface Gravity        &   $\log g = 8.5\pm 0.1$~(cgs)  \\*
WD Radius              &   $R_{\rm WD} = 0.01\pm 0.001~R_\odot$   \\*
WD Luminosity          &   $L_{\rm WD} = 9\pm 3~10^3~L_\odot$   \\*
WD Mass                &   $M_{\rm WD} = 1.3\pm 0.3~M_\odot$   \\*
\enddata
\end{deluxetable}




\begin{thebibliography}{}
\bibitem[Alcock et al.(1997)]{alcock97} Alcock, C., Allsman, R. A., Alves, D., et al. 1997,
   \mnras, 286, 483
\bibitem[Alcock et al.(2004)]{alcock04} Alcock, C., Alves, D. R., Axelrod, T. S., et al. 2004,
   \aj, 127, 334
\bibitem[Aldrovandi \& P\'equignot(1973)]{aldrovandi73} Aldrovandi, S. M. V., \&
   P\'equignot, D. 1973, \aap, 25, 137
\bibitem[Arnaud \& Raymond(1992)]{arnaud92} Arnaud, M., \& Raymond, J. 1992, \apj, 398, 394
\bibitem[Arnett(1969)]{arnett69} Arnett, D. W. 1969, \apss, 5, 180
\bibitem[Balucinska-Church \& McCammon(1992)]{BCMC92} Balucinska-Church, M., \& McCammon, D. 1992,
   \apj, 400, 699
\bibitem[Butler et al.(1993)]{ADOC19} Butler, K., Mendoza, C., \& Zeippen, C. J. 1993, 
    J. Phys. B, 26, 4409
\bibitem[Cassisi et al.(1998)]{cassisi98} Cassisi, S., Iben, I. Jr., \& Tornamb\`e, A. 1998,
   \apj, 496, 376
\bibitem[Clausen et al.(2003)]{clausen03} Clausen, J. V., Storm, J., Larsen, S. S., \&
   Gim\'enez, A. 2003, \aap, 402, 509
\bibitem[Cowley et al.(1984)]{cowley84} Cowley, A. P., Crampton, D., Hutchings, J. B., et al. 1984,
   \apj, 286, 196
\bibitem[Dreizler \& Werner(1993)]{dreizler93} Dreizler, S., \& Werner, K. 1993, \aap, 278, 199
\bibitem[Fernley et al.(1987)]{ADOC7} Fernley, J. A., Taylor, K. T., \& Seaton, M. J. 1987,
   J. Phys. B, 20, 6457
\bibitem[G\"{a}nsicke et al.(1998)]{gansicke98} G\"{a}nsicke, B. T., van Teeseling, A., 
   Beuermann, K., \& de Martino, D. 1998, \aap, 333, 163
\bibitem[Greiner et al.(1991)]{greiner91} Greiner, J., Hasinger, G., \& Kahabka, P. 1991, \aap,
   246, L17
\bibitem[Greiner \& Di Stefano(2002)]{greiner02} Greiner, J., \&  Di Stefano, R.  2002, \aap,
   387, 944 
\bibitem[Greiner et al.(2004)]{greiner04} Greiner, J., Iyudin, A., Jimenez-Garate, M., et al. 2004,
   to appear in Compact Binaries in the Galaxy and Beyond, astro-ph/0403426
\bibitem[Grevesse \& Sauval(1998)]{grevesse98} Grevesse, N., \& Sauval, A. J. 1998, Sp. Sci. Rev.,
   85, 161
\bibitem[Hachisu et al.(1996)]{hachisu96} Hachisu, I., Kato, M., \& Nomoto, K. 1996, \apj,
   470, L97
\bibitem[Hachisu et al.(1999)]{hachisu99} Hachisu, I., Kato, M., Nomoto, K., \& Umeda, H. 1999,
   \apj, 519, 314
\bibitem[Hamada \& Salpeter(1961)]{hamada61} Hamada, T., \& Salpeter, E. E. 1961, \apj, 134, 683
\bibitem[Hamuy et al.(2003)]{hamuy03} Hamuy, M., Phillips, M. M., Suntzeff, N. B., et al. 2003,
    Nature, 424, 651
\bibitem[Hartmann \& Heise(1997)]{hartmann97} Hartmann, H. W., \& Heise, J. 1997, \aap, 322, 591
\bibitem[Heise et al.(1994)]{heise94} Heise, J., van Teeseling, A., \& Kahabka, P. 1994,
   \aap, 288, L45
\bibitem[Hoyle \& Fowler(1960)]{hoyle60} Hoyle, F., \& Fowler, W. A. 1960, \apj, 132, 565
\bibitem[Hubeny et al.(2001)]{AGN4} Hubeny, I., Blaes, O., Krolik, J. H., \&
   Algol, E. 2001, \apj, 559, 680
\bibitem[Hubeny \& Lanz(1995)]{NLTE1} Hubeny, I., \& Lanz, T. 1995, \apj,
   439, 875
\bibitem[Iben(1982)]{iben82} Iben, I. Jr. 1982, \apj, 259, 244
\bibitem[Iben \& Tutukov(1984)]{iben84} Iben, I. Jr., \& Tutukov, A. V. 1984, \apjs, 54, 335
\bibitem[Ivanova \& Taam(2004)]{ivanova04} Ivanova, N., \& Taam, R. E. 2004, \apj, 601, 1058
\bibitem[Kahabka et al.(1996)]{kahabka96} Kahabka, P., Haberl, F., Parmar, A.~N., \& Greiner, J.
   1996, \iaucirc, 6467, 2 
\bibitem[Kahabka \& van den Heuvel(1997)]{kahabka97} Kahabka, P., \& van den Heuvel, E. P. J. 1997,
    \araa, 35, 69
\bibitem[Kahabka(1998)]{kahabka98} Kahabka, P. 1998, \aap, 331, 328
\bibitem[Kallman(2000)]{kallman00} Kallman, T. R. 2000, XSTAR: A Spectral Analysis Tool,
   User's Guide Version 2.0 (Greenbelt: NASA Goddard Space Flight Center)
\bibitem[Kato(1997)]{kato97} Kato, M. 1997, \apjs, 113, 121
\bibitem[Kraft et al.(1991)]{kraft91} Kraft, R.~P., Burrows, D.~N., \& Nousek, J.~A. 1991, \apj,
   374, 344 
\bibitem[Lanz \& Hubeny(1995)]{NLTE2} Lanz, T., \& Hubeny, I. 1995, \apj, 439, 905
\bibitem[Lanz \& Hubeny(2003)]{OS02} Lanz, T., \& Hubeny, I. 2003, \apjs, 146, 417
\bibitem[Livio \& Riess(2003)]{livio03} Livio, M., \& Riess, A. G. 2003, \apj, 594, L93
\bibitem[Luo \& Pradhan(1989)]{ADOC11} Luo, D., \& Pradhan, A. K. 1989, J. Phys. B, 22, 3377
\bibitem[Long et al.(1981)]{long81} Long, K. S., Helfand, D. J., \& Grabelsky, D. A. 1981,
   \apj, 248, 925
\bibitem[Mihalas(1978)]{SA78} Mihalas, D. 1978, Stellar Atmospheres, 2nd Ed.,
    (San Francisco: Freeman)
\bibitem[Napiwotzki et al.(2002)]{napiwotzki02} Napiwotzki, R., Koester, D., Nelemans, G., et al.
    2002, \aap, 386, 957
\bibitem[Napiwotzki et al.(2003)]{napiwotzki03} Napiwotzki, R., Christlieb, N., Drechsel, H., et al.
    2003, ESO Messenger, 112, 25
\bibitem[Nomoto et al.(1979)]{nomoto79} Nomoto, K., Nariai, K., \& Sugimoto, D. 1979,
   \pasj, 31, 287
\bibitem[Nomoto \& Iben(1985)]{nomoto85} Nomoto, K., \& Iben, I. Jr. 1985, \apj, 297, 531
\bibitem[Nussbaumer \& Storey(1983)]{nussbaumer83} Nussbaumer, H., \& Storey, P. J. 1983,
   \aap, 126, 75
\bibitem[OP(1995)]{IOP95} The Opacity Project Team, 1995, The Opacity Project,
    Vol. 1 (Bristol, UK: Inst. of Physics Publications)
\bibitem[OP(1997)]{IOP97} The Opacity Project Team, 1997, The Opacity Project,
    Vol. 2 (Bristol, UK: Inst. of Physics Publications)
\bibitem[Paerels et al.(2001)]{paerels01} Paerels, F., Rasmussen, A. P., Hartmann, H. W., 
   Heise, J., Brinkman, A. C., de Vries, C. P., \& den Herder, J. W. 2001, \aap, 365, L308
\bibitem[Parmar et al.(1998)]{parmar98} Parmar, A. N., Kahabka, P., Hartmann, H. W., 
   Heise, J., \& Taylor, B. G. 1998, \aap, 332, 199
\bibitem[Peach et al.(1988)]{ADOC9} Peach, G., Saraph, H. E., \&
    Seaton, M. J. 1988, J. Phys. B, 21, 3669
\bibitem[Rappaport et al.(1994)]{rappaport94} Rappaport, S., Di Stefano, R., \& Smith, J. D. 1994,
    \apj, 426, 692
\bibitem[Rolleston et al.(2002)]{rolleston02} Rolleston, W. R. J., Trundle, C., \& Dufton, P. L. 2002,
    \aap, 396, 53
\bibitem[Saffer et al.(1998)]{saffer98} Saffer, R. A., Livio, M., \& Yungelson, L. R.
    1998, \apj, 502, 394
\bibitem[Saio \& Nomoto(1985)]{saio85} Saio, H., \& Nomoto, K. 1985, \aap, 150, L21
\bibitem[Smale et al.(1988)]{smale88} Smale, A. P., Corbet, R. H. D., Charles, P. A., et al. 1988,
    \mnras, 233, 51
\bibitem[Tr\"umper et al.(1991)]{trumper91} Tr\"umper, J., Hasinger, G., Aschenbach, B., et al. 1991,
    Nature, 349, 579 
\bibitem[Tully et al.(1990)]{ADOC14} Tully, J. A., Seaton, M. J., \& Berrington, K. A. 1990, 
    J. Phys. B, 23, 3811
\bibitem[van den Heuvel et al.(1992)]{vdH92} van den Heuvel, E. P. J., Bhattacharya, D.,
   Nomoto, K., \& Rappaport, S. A. 1992, \aap, 262, 97
\bibitem[Webbink(1984)]{webbink84} Webbink, R. F. 1984, \apj, 277, 355
\bibitem[Weidemann(1987)]{weidemann87} Weidemann, V. 1987, \aap, 188, 74
\bibitem[Whelan \& Iben(1973)]{whelan73} Whelan, J., \& Iben, I. Jr. 1973, \apj, 186, 1007
\bibitem[Yoon \& Langer(2003)]{yoon03} Yoon, S-C., \& Langer, N. 2003, \aap, 412, L53
\bibitem[Yoon et al.(2004)]{yoon04} Yoon, S-C., Langer, N., \& Scheithauer, S. 2004, \aap, 425, 217
\end{thebibliography}
\end{document}